\documentclass[sort&compress,5p]{elsarticle}
\usepackage{graphicx,natbib,amssymb}
\usepackage[pdftex,colorlinks,citecolor=blue,bookmarks]{hyperref}
\usepackage{bm}
\usepackage{amsmath}
\begin{document}

\title{Neutrinoless $\beta\beta$ decay nuclear matrix elements in an isotopic chain}
\author{Tom\'as R. Rodr\'iguez}
\address{Institut f\"ur Kernphysik, Technische Universit\"at Darmstadt, Schlossgartenstr. 2, D-64289 Darmstadt, Germany}
\ead{tfrutos@theorie.ikp.physik.tu-darmstadt.de}
\author{Gabriel Mart\'inez-Pinedo}
\address{Institut f\"ur Kernphysik, Technische Universit\"at Darmstadt, Schlossgartenstr. 2, D-64289 Darmstadt, Germany}
\address{GSI Helmholtzzentrum
  f\"ur Schwerionenforschung, Plankstr. 1, D-64291 Darmstadt, Germany}
\begin{abstract}
  We analyze nuclear matrix elements (NME) of neutrinoless double beta
  decay calculated for the Cadmium isotopes. Energy density functional
  methods including beyond mean field effects such as symmetry
  restoration and shape mixing are used. Strong shell effects are
  found associated to the underlying nuclear structure of the initial
  and final nuclei. Furthermore, we show that NME for two-neutrino
  double beta decay evaluated in the closure approximation,
  $M^{2\nu}_{\mathrm{cl}}$, display a constant proportionality with
  respect to the Gamow-Teller part of the neutrinoless NME,
  $M^{0\nu}_{\mathrm{GT}}$. This opens the possibility of determining
  the $M^{0\nu}_{\mathrm{GT}}$ matrix elements from $\beta^{\mp}$
  Gamow-Teller strength functions. Finally, the interconnected role of deformation,
  pairing, configuration mixing and shell effects in the NMEs is
  discussed.
\end{abstract}
\begin{keyword}
Neutrinoless double beta decay, Nuclear matrix elements, Energy density functional methods.
\PACS 21.60.Jz, 23.40.-s, 23.40.Hc
\maketitle
\end{keyword}
\section{Introduction}~\label{intro}
Neutrinoless double beta decay ($0\nu\beta\beta$) is the most
promising process to disentangle the Majorana nature of the neutrino,
its effective mass and the mass hierarchy~\cite{RMP_80_481_2008}. In
this process, an even-even nucleus can not energetically decay into
the odd-odd neighbor but it is allowed into the even-even nucleus with
two protons more and two neutron less. Only few candidates which
fulfill this requirement are found along the nuclear chart in the
valley of the stability. Contrary to the two-neutrino double beta
decay mode ($2\nu\beta\beta$), where two electrons and two Dirac or
Majorana neutrinos are emitted in the final state, the
$0\nu\beta\beta$ mode proceeds by exchanging a Majorana neutrino and
only two electrons are emitted at a sharp energy equal to the
$Q$-value. $2\nu\beta\beta$ decay has been observed in
several isotopes with half-lives $~10^{19-21}$ years. However, except
to the controversial claim of the Heidelberg-Moscow
experiment~\cite{PLB_586_198_2004}, the neutrinoless mode has not been
measured yet. Nowadays, several experiments devoted to detect this
decay mode are presently active or in an advance development
phase~\cite{PPNP_64_249_2010,PAN_74_603_2011}. If this process is
finally detected, the precise determination of the neutrino effective
mass depends on the value of the nuclear matrix element (NME). The
half-life of this decay, in the so-called light neutrino exchange
mechanism, can be written as~\cite{RMP_80_481_2008}:
\begin{equation}
\left[T_{1/2}^{0\nu}(0_{i}^{+}\rightarrow0_{f}^{+})\right]^{-1}=G_{0\nu}|M^{0\nu}|^{2}\left(\frac{\langle
    m_{\beta\beta}\rangle}{m_{e}}\right)^{2} 
\label{half_life}
\end{equation}
where $G_{0\nu}$ is a well-known kinematic phase space
factor~\cite{PRC_85_034316_2012}, $M^{0\nu}$ is the NME, $m_{e}$ is
the electron mass and $\langle m_{\beta\beta}\rangle$ is the effective
Majorana neutrino mass.  Several nuclear structure methods have been
used so far to calculate these NMEs for the most promising candidates,
namely, large scale shell model
(SM)~\cite{PRL_100_052503_2008,NPA_818_139_2009}, quasiparticle random
phase approximation
(QRPA)~\cite{PRC_60_055502_1999,PRC_77_045503_2008,PRC_75_051303_2007},
interacting boson model (IBM)~\cite{PRC_79_044301_2009}, projected
Hartree-Fock-Bogoliubov (PHFB)~\cite{PRC_78_054302_2008} and energy
density functional methods
(EDF)~\cite{PRL_105_252503_2010,PPNP_66_436_2011}. In the most recent
calculations, the spread in the value of NMEs between different models
for a specific candidate is about a factor two. Therefore, it is
necessary to study in more detail the NMEs in order to better
constraint their values and understand possible relationships
  between the magnitude of the matrix element and the nuclear
  structure of parent and daughter states.  In this Letter we use
energy density functional methods including beyond mean field effects
to analyze the NMEs for the decay of Cadmium to Tin isotopes.  All of
these decays except the one of $^{116}$Cd are not physically possible
because of the $Q$-value or because they are strongly hindered by
single $\beta^{\pm}$ decays. However, these virtual
$0\nu\beta\beta$ decays provide useful information about the
dependence of the NMEs with the underlying nuclear structure of the
mother and daughter nuclei. Therefore, shell effects or the role of
deformation and pairing can be studied in a systematic and
controllable manner. The paper is organized as follows. The
theoretical formalism is described in Sec.~\ref{theo_frame}. Then, the
results obtained are reported in Sec.~\ref{results}. Finally, a
summary of the main conclusions is given in Sec.~\ref{summary}.
\section{Theoretical framework}~\label{theo_frame}
We now describe the main aspects of the generator coordinate method
(GCM) used to compute the NMEs with energy density functional
methods. In this framework, we determine the initial ($i$) and final
($f$) states as linear combinations of particle number and angular
momentum projected Hartree-Fock-Bogoliubov (HFB) mean-field wave
functions~\cite{RingSchuck,RMP_75_121_2003,NPA_696_467_2001,NPA_709_467_2001,PRL_99_062501_2007}:
\begin{equation}
|I^{+\sigma}_{i/f}\rangle=\sum_{\beta_{2}}g^{I\sigma}_{i/f}(\beta_{2})|\Psi^{I}_{i/f}(\beta_{2})\rangle
\label{GCM_WF}
\end{equation}
where $I$ is the angular momentum, $\beta_{2}$ are intrinsic axial
quadrupole deformations, $g^{I\sigma}_{i/f}(\beta_{2})$ are the
coefficients found by solving the associated Hill-Wheeler-Griffin
(HWG) equations~\cite{RingSchuck}:
\begin{eqnarray}
\sum_{\beta'_{2}}\left(\langle\Psi^{I}_{i/f}(\beta_{2})|\hat{H}|\Psi^{I}_{i/f}(\beta'_{2})\rangle\right.\nonumber\\-E^{I\sigma}_{i/f}\left.\langle\Psi^{I}_{i/f}(\beta_{2})|\Psi^{I}_{i/f}(\beta'_{2})\rangle\right)g^{I\sigma}_{i/f}(\beta'_{2})=0
\label{HWG_eq1}
\end{eqnarray}
Here, $E^{I\sigma}_{i/f}$ are the energy spectra of the initial/final
nuclei and the projected wave functions are defined as:
\begin{equation}
|\Psi^{I}_{i/f}(\beta_{2})\rangle=P^{N_{i/f}}P^{Z_{i/f}}P^{I}|\phi(\beta_{2})\rangle
\label{PROJ_WF}
\end{equation}
with $P^{N}$, $P^{Z}$, $P^{I}$ being the projection operators onto
good number of neutrons, protons and angular momentum,
respectively. Furthermore, for each deformation $\beta_{2}$, the
HFB-type states $|\phi(\beta_{2})\rangle$ are found by minimizing the
particle number projected energy with constraints in the mean value of
the axial quadrupole moment operator $\hat{Q}_{20}$, i.e.,
$\delta(E^{'NZ}(|\phi(\beta_{2})\rangle))=0$
with~\cite{NPA_696_467_2001}:
\begin{equation}
E^{'NZ}=\frac{\langle\phi(\beta_{2})|\hat{H}P^{N}P^{Z}|\phi(\beta_{2})\rangle}{\langle\phi(\beta_{2})|\phi(\beta_{2})\rangle}-\lambda_{\beta_{2}}\langle\phi(\beta_{2})|\hat{Q}_{20}|\phi(\beta_{2})\rangle
\end{equation}
The Lagrange multiplier $\lambda_{\beta_{2}}$ ensures the condition:
\begin{equation}
\langle\phi(\beta_{2})|\hat{Q}_{20}|\phi(\beta_{2})\rangle=\frac{\beta_{2}3r_{0}^{2}A^{5/3}}{\sqrt{20\pi}}
\end{equation}
with $r_{0}=1.2$ fm and $A$ the mass number.  We use the same
interaction both in the determination of the intrinsic states and in
the HWG diagonalization, in our case the Gogny D1S
force~\cite{NPA_428_23_1984}.  To solve the HWG equations we transform
first Eq.~\ref{GCM_WF} in terms of an orthonormal basis:
\begin{equation}
|\Lambda^{I}_{i/f}\rangle=\sum_{\beta_{2}}\frac{u^{I}_{\Lambda_{i/f}}(\beta_{2})}{\sqrt{n^{I}_{\Lambda_{i/f}}}}|\Psi^{I}_{i/f}(\beta_{2})\rangle;\,\,n^{I}_{\Lambda_{i/f}}>\varepsilon
\end{equation}
where $u^{I}_{\Lambda_{i/f}}(\beta_{2})$ and $n^{I}_{\Lambda_{i/f}}$ are the eigenvectors and eigenvalues -greater than a small value $\varepsilon\sim10^{-5}$ chosen to remove linear dependence of the states- of the norm overlap matrix:
\begin{equation}
\sum_{\beta'_{2}}\langle\Psi^{I}_{i/f}(\beta_{2})|\Psi^{I}_{i/f}(\beta'_{2})\rangle u^{I}_{\Lambda_{i/f}}(\beta'_{2})=n^{I}_{\Lambda_{i/f}} u^{I}_{\Lambda_{i/f}}(\beta_{2})
\end{equation}
Then, the states given in Eq.~\ref{PROJ_WF} can be expressed as $|I^{+\sigma}_{i/f}\rangle=\sum_{\Lambda_{i/f}}G^{I\sigma}_{\Lambda_{i/f}}|\Lambda^{I}_{i/f}\rangle$ and the HGW equations (Eq.~\ref{HWG_eq1}) read as:
\begin{equation}
\sum_{\Lambda'_{i/f}}\langle\Lambda^{I}_{i/f}|\hat{H}|\Lambda^{'I}_{i/f}\rangle G^{I\sigma}_{\Lambda'_{i/f}}=E^{I\sigma}_{i/f}G^{I\sigma}_{\Lambda_{i/f}}
\label{HWG_eq2}
\end{equation}
The eigenvectors $G^{I\sigma}_{\Lambda_{i/f}}$ are used to define the
relative weight of each deformation $\beta_{2}$ in the GCM wave
function, the so-called collective wave functions~\cite{RingSchuck}:
\begin{equation}
F^{I\sigma}_{i/f}(\beta_{2})=\sum_{\Lambda_{i/f}}G^{I\sigma}_{\Lambda_{i/f}}u^{I}_{\Lambda_{i/f}}(\beta_{2})
\label{coll_wf}
\end{equation} 
In addition, any expectation value between GCM wave functions
(energies, radii, transitions, etc.) can be expressed in terms of
these coefficients, as we will see in the case of $0\nu\beta\beta$
NMEs.

It is important to underline that, in this framework, particle number
and rotational symmetries are conserved by projecting the intrinsic
HFB wave functions and, in addition, shape mixing is naturally
included and self-consistently determined by solving the HWG
equations. On the other hand, the main restrictions of the model come
from the symmetries imposed to the HFB-type states
$|\phi(\beta_{2})\rangle$. They are tensorial products of proton and
neutron wave functions (no isospin mixing), and are also parity and
axially symmetric. Work is in progress to check the effect of these
approximations in the NMEs. Finally, a large configuration space
including eleven major harmonic oscillator shells and sets of 56
states with different deformation ranging from
$\beta_{2}\in[-0.5,0.6]$ are used in this work.

Once the initial and final states are found, we proceed to evaluate
the NMEs as the sum of Fermi (F) and Gamow-Teller (GT)
terms~\cite{RMP_80_481_2008}(tensor contribution is neglected in this
work~\cite{NPA_818_139_2009,PRC_75_051303_2007}):
 
\begin{equation}
M^{0\nu}=-\left(\frac{g_{V}}{g_{A}}\right)^{2}M^{0\nu}_{F} +M^{0\nu}_{GT}
\label{MMM}
\end{equation}
with $g_{V}=1$ and $g_{A}=1.25$ being the vector and axial coupling
constants.  First, we use the closure approximation to by-pass the
calculation of the odd-odd intermediate nucleus. This approximation is
expected to be a good one in the $0\nu\beta\beta$
case~\cite{RMP_80_481_2008,PRC_83_015502_2011}. Then, the NMEs can be determined as the
expectation value of two-body operators between the initial and final
states:

\begin{equation}
M^{0\nu}_{F/GT}=\langle
0^{+}_{f}|\hat{M}^{0\nu}_{F/GT}|0^{+}_{i}\rangle\label{NME_EV}
\end{equation} 
with:
\begin{eqnarray}
  \label{eq:1}
  \hat{M}^{0\nu}_{F}& = &\left(\frac{g_{A}}{g_{V}}\right)^{2}\sum_{i<j} \hat{V}_{F}(r_{ij})
  \hat{\tau}^{(i)}_{-}\hat{\tau}^{(j)}_{-}, \\
  \hat{M}^{0\nu}_{GT} & = & \sum_{i<j} \hat{V}_{GT}(r_{ij})
  (\hat{\bm{\sigma}}^{(i)} \cdot
  \hat{\bm{\sigma}}^{(j)})\hat{\tau}^{(i)}_{-}\hat{\tau}^{(j)}_{-}     
\end{eqnarray}
In these expressions, $\hat{\tau}_{-}$ is the isospin ladder operator
that changes neutrons into protons and $\hat{\bm{\sigma}}$ are the
Pauli matrices acting on the spin part of the wave
functions. The so-called neutrino potentials $\hat{V}_{F/GT}$ depend on the relative distance between two nucleons. These
potentials take into account nucleon finite
size corrections~ and higher order currents~\cite{PRC_60_055502_1999}
(see Refs.~\cite{PRC_60_055502_1999,NPA_818_139_2009} for detailed
expressions). Before including short-range correlations within the UCOM framework~\cite{NPA_632_61_1998,PRC_77_045503_2008} the neutrino
potentials can be expressed as an integral in the momentum transferred
$q$ of the Fermi or Gamow-Teller form factor $h_{F/GT}(q)$ weighted by
the spherical Bessel function
$j_{0}(qr)$~\cite{PRC_60_055502_1999,NPA_818_139_2009}:
\begin{equation}
V_{F/GT}(r)=\frac{2}{\pi}\frac{r_{0}A^{1/3}}{g^{2}_{A}}\int_{0}^{\infty}j_{0}(qr)\frac{h_{F/GT}(q)}{q+\mu}qdq 
\label{neutrino_pot}
\end{equation}
For the shake of simplicity we take the same
value of the closure energy $\mu=10.22$ MeV for the whole isotopic
chain. This value is the one used for computing the NME of $^{116}$Cd
decay in Ref.~\cite{PRL_105_252503_2010}. 

We now calculate the NMEs (Eq.~\ref{NME_EV}) with the EDF formalism
described above. We take the ground states given by the lowest energy
solution of the HWG equation with $(I=0,\sigma=1)$ and evaluate: 
\begin{eqnarray}
M^{0\nu}_{F/GT}=\sum_{\begin{subarray}\.\Lambda_{f}\Lambda_{i}\\
\beta_{2}\beta'_{2}\end{subarray}}G^{*0,1}_{\Lambda_{f}}\frac{u^{*0}_{\Lambda_{f}}(\beta_{2})}{\sqrt{n^{0}_{\Lambda_{f}}}}\frac{u^{0}_{\Lambda_{i}}(\beta'_{2})}{\sqrt{n^{0}_{\Lambda_{i}}}}G^{0,1}_{\Lambda_{i}}\nonumber\\\bar{M}^{0\nu}_{F/GT}(\beta_{2},\beta'_{2})
\label{NME_final}
\end{eqnarray}
where
$\bar{M}^{0\nu}_{F/GT}(\beta_{2},\beta'_{2})=\langle\Psi^{0}_{f}(\beta_{2})|\hat{M}^{0\nu}_{F/GT}|\Psi^{0}_{i}(\beta'_{2})\rangle$. The
last expression, properly normalized, gives the explicit dependence of
the NMEs with the quadrupole deformation of the mother and daughter
nuclei: 
\begin{equation}
  M^{0\nu}_{F/GT}(\beta_{2},\beta'_{2})=\frac{\bar{M}^{0\nu}_{F/GT}(\beta_{2},\beta'_{2})}{\sqrt{\langle\Psi^{0}_{f}(\beta_{2})|\Psi^{0}_{f}(\beta_{2})\rangle\langle\Psi^{0}_{i}(\beta'_{2})|\Psi^{0}_{i}(\beta'_{2})\rangle}} 
  \label{NME_beta}
\end{equation}
The final value of the NME can be approximately interpreted as the
convolution of the collective wave functions of initial and final
states (Eq.~\ref{coll_wf}) with the intensity of the Fermi and GT NMEs
as a function of the deformation given in Eq.~\ref{NME_beta}.
\section{Results}~\label{results}
\begin{figure}[t]
\begin{center}
  \includegraphics[width=\columnwidth]{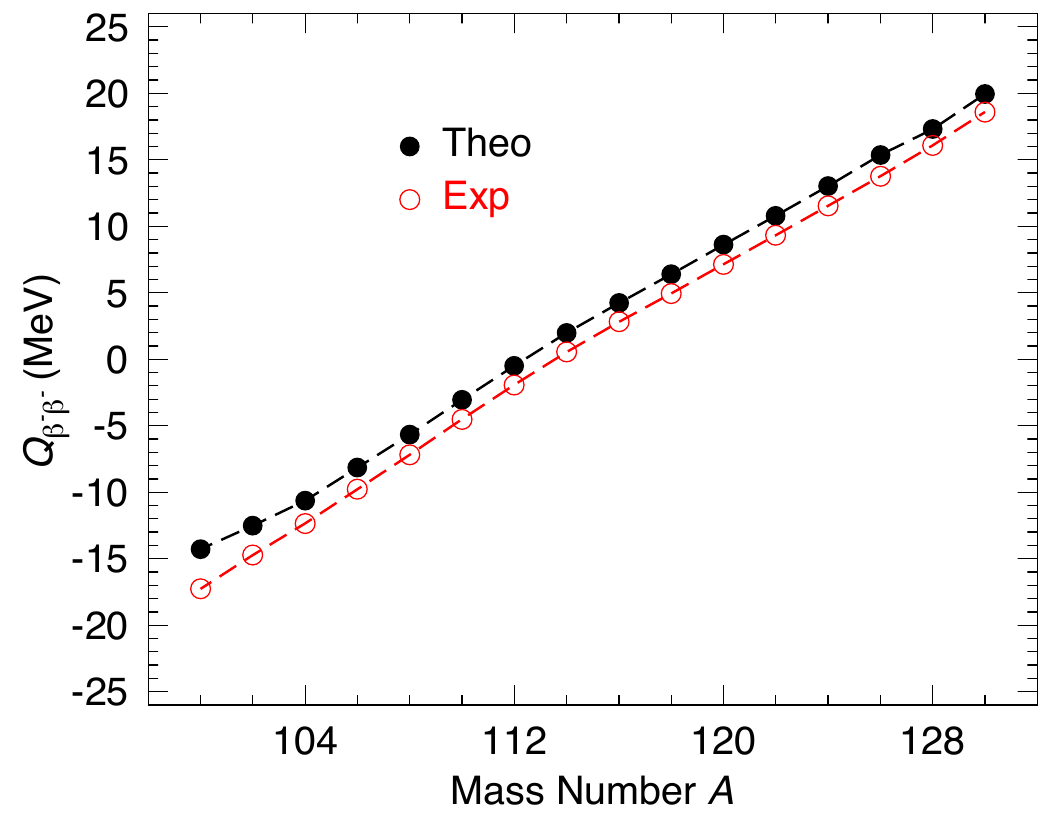}
\end{center}
\caption{(color online) Experimental~\cite{NPA_729_337_2003} and
  theoretical $Q_{\beta^{-}\beta^{-}}$-value for cadmium isotopes as a
  function of the mass number $A$.}\label{Fig1} 
\end{figure}
We now show the results of the $0\nu\beta\beta$ NMEs in the cadmium
isotopic chain $^{98\mathrm{-}132}$Cd. This set of nuclei covers the
shells from $N=50$ to $N=82$ magic numbers both in the mother (Cd) and
daughter (Sn) isotopes. 
We first describe the ground states of these nuclei that are the ones
involved in the $0\nu\beta\beta$ decay. In Fig.~\ref{Fig1} the
experimental~\cite{NPA_729_337_2003} and theoretical
$Q_{\beta^{-}\beta^{-}}$- values are plotted. We observe that
experimental and theoretical data almost run parallel each other,
increasing the value continuously with increasing the mass number from
negative to positive values. Additionally, a quantitative agreement is
found within the limit of accuracy of the Gogny D1S force, which is
approximately 3.4~MeV~\cite{EPJA_33_237_2007}. Nevertheless, the
theoretical result overestimates the $Q_{\beta^{-}\beta^{-}}$-value
about $\sim1.4$ MeV. It is important to point out that no tuning
of the parameters of the interaction has been made. 

\begin{figure}[t]
  \begin{center}
    \includegraphics[width=0.9\columnwidth]{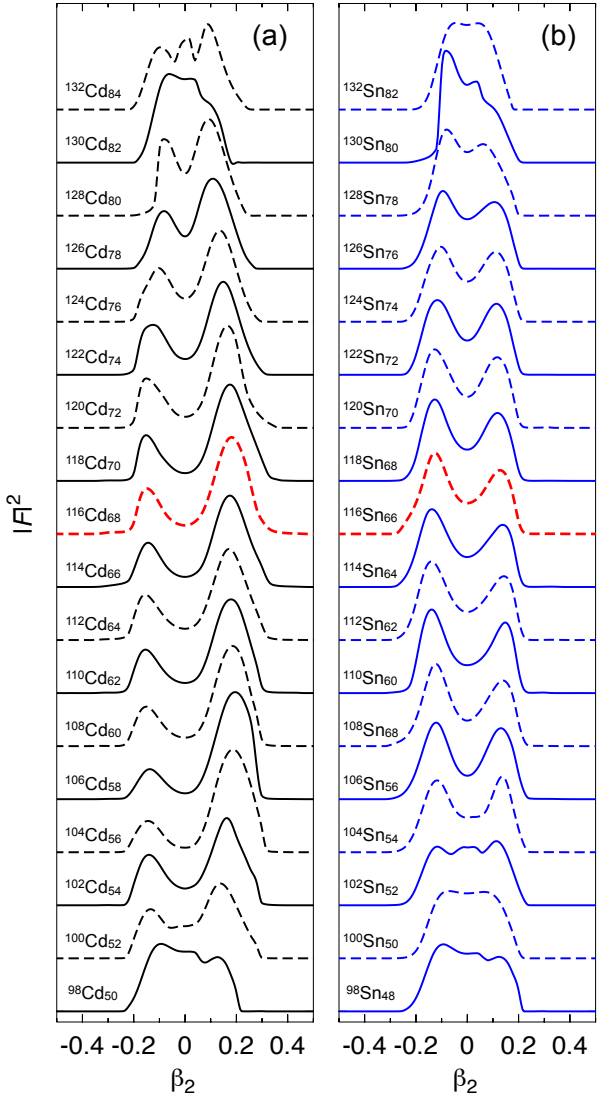}
  \end{center}
  \caption{(color online)Ground state $0^{+}$ collective wave
    functions for (a) cadmium and (b) tin isotopes calculated with the
    Gogny D1S EDF. $^{116}$Cd ($^{116}$Sn) double beta emitter is
    represented in red.}\label{Fig2} 
\end{figure}

In Fig.~\ref{Fig2} we represent the evolution from shell to shell of
the ground state collective wave functions (Eq.~\ref{coll_wf}) for Cd
-Fig.~\ref{Fig2}(a)- and Sn -Fig.~\ref{Fig2}(b)- isotopes. First of
all, we observe that the distributions are enclosed around
$\beta_{2}\sim\left[-0.2,0.2\right]$ for all nuclei, i.e., these are
not very deformed systems. In addition, most of the collective wave
functions present two maxima, one oblate and one prolate, with a
minimum in the spherical point. This is not the case for the magic or
their nearest neighbor nuclei ($^{98,130\mathrm{-}132}$Cd and
$^{98\mathrm{-}102,130\mathrm{-}132}$Sn) whose collective wave functions also peak in
the spherical point. For the Cd chain (Fig.~\ref{Fig2}(a)) prolate
maxima are higher than the oblate ones while a more symmetric
distribution around $\beta_{2}=0$ is obtained for the Sn isotopes. In
addition, mirror nuclei $^{98}$Cd and $^{98}$Sn have almost identical
collective wave functions as one could expect. 

\begin{figure}[t]
\begin{center}
  \includegraphics[width=\columnwidth]{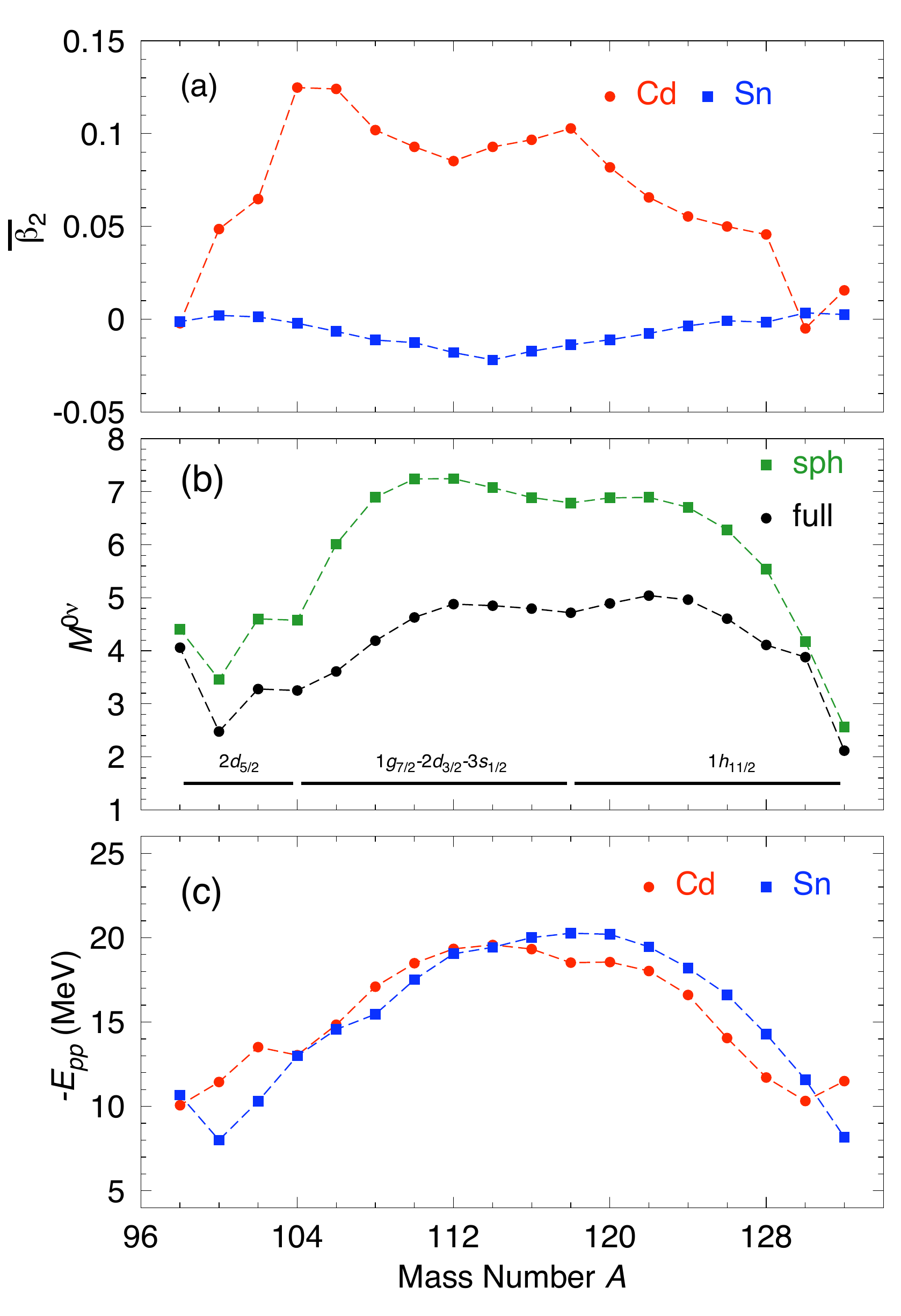}
\end{center}
\caption{(color online) (a) Mean value of the quadrupole deformation
  $\beta_{2}$ for the collective wave functions given in
  Fig.~\ref{Fig2}. (b) Nuclear matrix elements for initial and final
  nuclei considered as spherical systems (filled squares) and
  including the full shape mixing (bullets). (c) Total pairing
  energies for Cd and Sn isotopes including shape configuration
  mixing.\label{Fig3}}
\end{figure}

A more quantitative analysis of the deformation can be done by
representing in Fig.~\ref{Fig3}(a) the mean value
$\bar{\beta}_{2}$ for the collective wave functions given
above. We see in Fig.~\ref{Fig3}(a) that tin isotopes are spherical or
very little oblate deformed in the mid-shell
($\bar{\beta}_{2}\sim-0.025$ for $^{114}$Sn) as it is expected
from its proton shell closure $Z=50$. On the other hand, the mean
deformation in cadmium isotopes is zero only in the shell closures
$N=50,82$ while the rest of nuclei are slightly prolate
deformed. Hence, we find two maxima at $^{104,106}$Cd and $^{118}$Cd
with $\bar{\beta}_{2}\sim0.12$ and $\bar{\beta}_{2}\sim0.10$
respectively and a local minimum at $^{112}$Cd
($\bar{\beta}_{2}\sim0.08$). 

The effect of the deformation of initial and final states and the role
played by pairing correlations on the NMEs for individual decays have
been already discussed extensively within the EDF
framework~\cite{PRL_105_252503_2010,PPNP_66_436_2011}, and, to a
lesser extent, within the shell
model~\cite{JPCS_267_012058_2011}, QRPA~\cite{PRC_83_034320_2011} and
IBM~\cite{PRC_79_044301_2009} approaches. On the one hand, a
strong suppression of the NMEs is found for decays between states with
different intrinsic deformations. Furthermore, shape mixing effects
tend to the reduce the values of the matrix elements calculated
assuming only spherical
symmetry~\cite{PPNP_66_436_2011,PRC_85_044310_2012}. On the other
hand, the NMEs are enhanced if a large amount of pairing correlations
are present, or in the shell model language, when the
  wave functions are dominated by generalize seniority zero
  components, in the mother and daughter
nuclei~\cite{PRL_105_252503_2010,PRL_100_052503_2008}. Let us study
now if these conclusions can be extended to the NMEs calculated in the
Cd isotopic chain where the number of neutrons changes smoothly within
a major shell.

In Fig.~\ref{Fig3}(b) we compare the values of the NMEs
  obtained from the full configuration mixing (Eq.~\ref{NME_final})
  with those assuming spherical shapes
  $M^{0\nu}(\beta_{2}=0,\beta'_{2}=0)$ (Eq.~\ref{NME_beta}) for mother
  and daughter nuclei. We observe that, except for the
transition between mirror nuclei $A=98$, $M^{0\nu}$ is small near the
neutron shell closures. Furthermore, the NMEs for the spherical shape
are always larger than the ones which include configuration mixing
although both curves show a similar structure. The reduction
obtained by shape mixing is smaller near the shell closures where the
relevant deformations explored by the collective wave functions are
precisely the spherical ones. From Fig~\ref{Fig3}(a) we can also see
qualitatively that the larger is the difference between the mean
quadrupole deformation of mother and daughter nuclei the larger is the
reduction of the NME with respect to its spherical value. This shows
that the correlation observed between the value of NME and differences
in nuclear deformation is a general feature of neutrinoless double
beta-decay and not a particular aspect of those beta-decays studied
so far~\cite{PRL_105_252503_2010}.

We now analyze the structure of the NMEs along the isotopic chain
(Fig.~\ref{Fig3}(b)). First, we observe a large value of the NME for
$A=98$ transition between mirror nuclei. The wave functions of the
initial and final states are practically identical (see
Fig.~\ref{Fig2}) and their overlap is therefore
maximized~\cite{JPCS_267_012058_2011}. Apart from this specific case,
the rest of the curve shows some structure that can be related to
neutron sub-shell closures in Cd isotopes. In
Ref.~\cite{PRC_79_044301_2009} shell effects are studied within a
generalized seniority scheme and it is observed that the lowest NME
values correspond to the shell closures and a maximum is found at the
mid-shell. The same behavior can be also seen in Fig.~\ref{Fig3}(b)
where three maxima are found at $A(N)=102(54),112(64)$ and $122(74)$
that can be related to the half-filling of
$2d_{5/2}$, $1g_{7/2}3s_{1/2}2d_{3/2}$ and $1h_{11/2}$
sub-shells. The local minimum found at $A=118$ would correspond to the
clossing of the $1g_{7/2}3s_{1/2}2d_{3/2}$ sub-shell.

These shell effects are also related to the content of pairing
correlations in initial and final nuclei as we can see in
Fig~\ref{Fig3}(c). In this figure we represent the pairing energy
($-E_{pp}$)~\cite{RingSchuck} evaluated with the GCM states given in
Eq.~\ref{GCM_WF}. In closed shell
nuclei, $N=50$ and $N=82$, pairing correlations are significantly
smaller than in open-shell nuclei, obtaining the parabolic shape given
in the figure. Nevertheless, we also find local minima in the Cd
isotopic chain at $A=104$ and $A=118$ related to the sub-shell
closures described above. Therefore, we obtain again a direct
generalization of a result already found in individual decays, namely,
the utter connection between pairing
correlations in the initial and final nuclei and the NMEs.

\begin{figure}[t]
\begin{center}
  \includegraphics[width=\columnwidth]{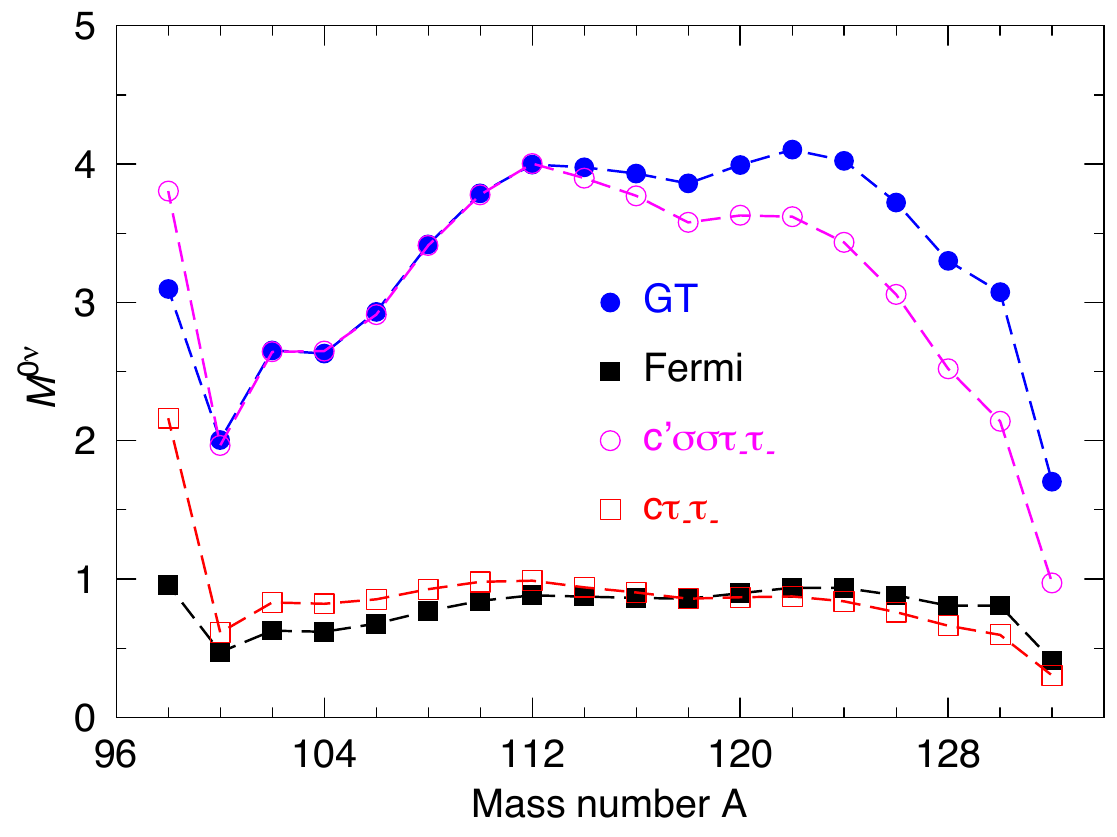}
\end{center}
\caption{(color online)  Nuclear matrix elements calculated with Fermi (squares), Gamow-Teller (bullets) and constant neutrino potentials $V_{F}(r)=c=2.0$ (boxes) and $V_{GT}(r)=c'=2.0$ (circles)\label{Fig4}}
\end{figure}

Finally we study the sensitivity of the NMEs to the form of the
neutrino potentials $V_{F/GT}(r)$ given in Eq.~\ref{neutrino_pot}. In
Fig.~\ref{Fig4} we compare the Fermi and GT components of $M^{0\nu}$
assuming a constant neutrino potential, i.e. independent of the
relative distance of the decaying neutrons, with those obtained using
the usual form~\eqref{neutrino_pot}. We take as 2.0 the constant value
of the neutrino potential as it reproduces both the Fermi and
Gamow-Teller NME computed using the full potential. The same pattern
is found for the NMEs calculated with a constant spatial dependence
and the ones given by the full neutrino potentials, i.e., similar
enhancement of $M^{0\nu}$ in mirror nuclei and shell effects both for
Fermi and Gamow-Teller parts. This shows that the overall behavior of
the NME is
quite insensitive to the details of the neutrino
potential. Nevertheless, the NMEs computed with regular neutrino
potentials are more symmetric with respect to the local minimum found
at $A=118$, while the ones computed with constant potentials are
smaller after this minimum. This fact is related to the filling of the
negative parity $1h_{11/2}$ sub-shell and could be associated to a
increased contribution of multipoles of higher order that the
Gamow-Teller, i.e. the only one present in the constant potential
case. 

This proportionality is important because the GT part calculated with
a constant potential is directly related to the $2\nu\beta\beta$
matrix element evaluated in the closure
approximation~\cite{PRC_83_015502_2011}:

\begin{equation}
  \label{eq:2}
     M^{2\nu}_{\mathrm{cl}} = \langle 0^+_f | \sum_{i<j} \hat{\bm{\sigma}}^{(i)} \cdot
  \hat{\bm{\sigma}}^{(j)} \hat{\tau}^{(i)}_{-}\hat{\tau}^{(j)}_{-} |
  0^+_i\rangle 
\end{equation}
Limits on this matrix element can be experimentally obtained from
charge exchange reactions (see~\cite{PRC_71_054313_2005} and
references therein) or assuming single-state dominance
hypothesis~\cite{AFA_80_9_1984}. Hence, it would help to constrain
experimentally the value of $M^{0\nu}_{GT}$. In this work we obtain
for $^{116}$Cd an unquenched value of $M^{2\nu}_{\mathrm{cl}}=1.885$
which is much larger than the one extracted from charge exchange
reactions~\cite{PRC_71_054313_2005}
$M^{2\nu}_{\mathrm{cl,cer}}=0.314$. However, the GT operator would require a
quenching factor to account for renormalization effects of the
weak-axial current. On the other hand, the experimental data reported
on~\cite{PRC_71_054313_2005} includes GT strength located at
excitation energies below 3~MeV in the intermedium nucleus. One expects
substantial contributions to the $M^{2\nu}_{\mathrm{cl}}$ from states in
the GT resonance at excitation energies around 10~MeV. Notice, that
these transitions are suppressed for the regular $M^{2\nu}$ matrix
element~\cite{PRC_83_015502_2011} due to the energy denominator but
not for the $M^{2\nu}_{\mathrm{cl}}$ matrix element. Based on QRPA
calculations ref.~\cite{PRC_83_015502_2011} shows that a converged
value of $M^{2\nu}_{cl}$ requires the inclusion of states up to 15~MeV
excitation energy in the intermedia nucleus. Further work is in
progress to determine the origin of this proportionality which has
also been observed, within the EDF framework, in the NMEs as a
function of deformation of individual decays~\cite{PPNP_66_436_2011},
and in shell model calculations~\cite{JPCS_267_012058_2011}, but not
found in QRPA calculations~\cite{PRC_83_015502_2011}. 
\section{Summary}~\label{summary}
In summary, we have studied the $0\nu\beta\beta$ nuclear matrix
elements in the cadmium isotopic chain with energy density functional
methods. It is shown that previous correlations found for individual
decays between the magnitude of the NME and pairing correlations,
deformation, shell effects and shape of the neutrino potential can be
extended to a whole isotopic chain. This shows that this correlations
are a particular feature of the NME that is independent of the
detailed structure of the nuclear wave functions. Large NMEs are
obtained if the pairing correlations in the initial and final states
are considerable and the difference in deformation is
small. Furthermore, the NMEs are very sensitive to the sub-shell
closures. In addition, the NME for mirror nuclei is enhanced due to
the large overlap between mother and daughter wave functions. Finally,
NMEs calculated neglecting the neutrino potential, i.e. the
two-neutrino double beta-decay matrix element evaluated in the closure
approximation, are found to exhibit a nearly constant proportionality to
the realistic ones. This opens the possibility of constraining the
$M^{0\nu}$ matrix element from the knowledge of the $\beta^{\mp}$
Gamow-Teller strength functions.

We thank J. Men\'endez and A. Poves for stimulating discussions.  
This work was partly supported by the Helmholtz International Center for
FAIR within the framework of the LOEWE program launched by the State of
Hesse, by the Deutsche Forschungsgemeinschaft through contract SFB~634 and by
the BMBF-Verbundforschungsprojekt number 06DA7047I.


\begin{thebibliography}{9}   
\bibitem{RMP_80_481_2008}
 F.~T. Avignone, S.~R. Elliot, J. Engel, Rev. Mod. Phys. 80, 481 (2008).
\bibitem{PLB_586_198_2004} 
H. V. Klapdor-Kleingrothaus \textit{et al.}, Phys. Lett. B 586, 198 (2004).
\bibitem{PPNP_64_249_2010} 
H. Ejiri, Prog. Part. Nucl. Phys. 64, 249 (2010).
\bibitem{PAN_74_603_2011}
A. S. Barabash, Phys. At. Nucl. 74, 603 (2011).
\bibitem{PRC_85_034316_2012}
J. Kotila and F. Iachello, Phys. Rev. C 85, 034316 (2012).
\bibitem{PRL_100_052503_2008} 
E. Caurier \textit{et al.}, Phys. Rev. Lett. 100, 052503 (2008). 
\bibitem{NPA_818_139_2009} 
J. Men\'endez \textit{et al.}, Nucl. Phys. A 818, 139 (2009). 
\bibitem{PRC_60_055502_1999} 
F. \u{S}imkovic \textit{et al.}, Phys. Rev. C 60, 055502 (1999). 
\bibitem{PRC_77_045503_2008} 
F. \u{S}imkovic \textit{et al.}, Phys. Rev. C 77, 045503 (2008). 
\bibitem{PRC_75_051303_2007} 
M. Kortelainen, J. Suhonen, Phys. Rev C 75, 051303(R) (2007).
\bibitem{PRC_79_044301_2009}
J. Barea and F. Iachello, Phys. Rev. C \textbf{79}, 044301 (2009).
\bibitem{PRC_78_054302_2008} 
K. Chaturvedi \textit{et al.}, Phys. Rev. C 78, 054302 (2008). 
\bibitem{PRL_105_252503_2010}
T. R. Rodr\'iguez and G. Martinez-Pinedo, Phys. Rev. Lett. 105, 252503 (2010).
\bibitem{PPNP_66_436_2011} 
T. R. Rodr\'iguez and G. Martinez-Pinedo, Prog. Part. Nucl. Phys. 66, 436 (2011).
\bibitem{RingSchuck} 
P. Ring, P. Schuck, \textit{The nuclear many body problem}, Springer-Verlag, Berlin, 1980. 
\bibitem{RMP_75_121_2003} 
M. Bender, P.-H. Heenen, P.-G. Reinhard, Rev. Mod. Phys. 75, 121 (2003). 
\bibitem{NPA_696_467_2001} 
M. Anguiano, J. L. Egido, L. M. Robledo, Nucl. Phys. A 696, 467 (2001). 
\bibitem{NPA_709_467_2001} 
R. R. Rodr\'iguez-Guzm\'an, J. L. Egido, L. M. Robledo, Nucl. Phys. A 709, 201 (2002). 
\bibitem{PRL_99_062501_2007} 
T.R. Rodr\'iguez, J.L. Egido, Phys. Rev. Lett. 99, 062501 (2007). 
\bibitem{NPA_428_23_1984}
J.~F.  Berger  et \textit{al.}, Nucl. Phys. A \textbf{428}, 23 (1984).
\bibitem{PRC_83_015502_2011}
F. \u{S}imkovic, R. Hod\'ak, A. Faessler, and P. Vogel, Phys. Rev. C 83, 015502 (2011).
\bibitem{NPA_632_61_1998}
H. Feldmeier, T. Neff, R. Roth, J. Schnack, Nucl. Phys. A \textbf{632}, 61 (1998).
\bibitem{NPA_729_337_2003}
G. Audi, A.H. Wapstra, C. Thibault, Nucl. Phys. A \textbf{729}, 337 (2003).
\bibitem{EPJA_33_237_2007}
S. Hilaire and M. Girod, Eur. Phys. J. A \textbf{33}, 237 (2007).
\bibitem{JPCS_267_012058_2011}
J. Men\'endez et \textit{al.} J. Phys.: Conf. Ser. 267, 012058 (2011).
\bibitem{PRC_83_034320_2011}
D.~-L. Fang, A. Faessler, V. Rodin, and F. \u{S}imkovic, Phys. Rev. C 83, 034320 (2011).
\bibitem{PRC_85_044310_2012}
T.R. Rodr\'iguez and Gabriel Mart\'inez-Pinedo, Phys. Rev. C 85, 044310 (2012).
\bibitem{PRC_71_054313_2005}
S. Rakers et \textit{al.}, Phys. Rev. C 71, 054313 (2005).
\bibitem{AFA_80_9_1984}
J. Abad, A. Morales, R. N\'u\~nez Lagos and A.F. Pacheco, An. Fis. A80, 9 (1984).
\end{thebibliography}
\end{document}